\documentclass[10pt, conference, letterpaper]{IEEEtran}
\usepackage[ruled,vlined,linesnumbered]{algorithm2e}
\usepackage{hhline}
\usepackage{amsmath,mathtools}
\usepackage{amsfonts,amssymb}
\usepackage{mathrsfs}
\usepackage{gensymb} 
\usepackage{graphicx} 
\usepackage{multirow}
\usepackage{enumitem,color}
\usepackage{algpseudocode}
\usepackage{makecell}
\usepackage{booktabs}
\usepackage{dsfont} 
\usepackage{caption}
\usepackage{subfigure} 
\usepackage{hyperref}
\usepackage{geometry}
\usepackage{titlesec}
\titlespacing*{\section}{1pt}{0.5ex}{0.5ex}
\titlespacing*{\subsection}{1pt}{0.5ex}{0.5ex}
\titlespacing*{\subsubsection}{1pt}{0.5ex}{0.5ex}

\begin{document}
%
\title{\huge Automatic Network Planning with Digital Radio Twin\vspace{-0.1in}}

\author{
\IEEEauthorblockN{
Xiaomeng Li\IEEEauthorrefmark{1}, 
Yuru Zhang\IEEEauthorrefmark{1}, 
Qiang Liu\IEEEauthorrefmark{1}, 
Mehmet Can Vuran\IEEEauthorrefmark{1}, \\
Nathan Huynh\IEEEauthorrefmark{1},
Li Zhao\IEEEauthorrefmark{1}, 
Mizan Rahman\IEEEauthorrefmark{2}, 
Eren Erman Ozguven\IEEEauthorrefmark{3} \\
\IEEEauthorrefmark{1} University of Nebraska-Lincoln,
\IEEEauthorrefmark{2} University of Alabama,
\IEEEauthorrefmark{3} Florida State University
\vspace{-0.2in}
}
}

\maketitle

\begin{abstract}
Network planning seeks to determine base station parameters that maximize coverage and capacity in cellular networks. However, achieving optimal planning remains challenging due to the diversity of deployment scenarios and the significant simulation-to-reality discrepancy.
In this paper, we propose \emph{AutoPlan}, a new automatic network planning framework by leveraging digital radio twin (DRT) techniques.
We derive the DRT by finetuning the parameters of building materials to reduce the sim-to-real discrepancy based on crowdsource real-world user data.
Leveraging the DRT, we design a Bayesian optimization based algorithm to optimize the deployment parameters of base stations efficiently.
Using the field measurement from Husker-Net, we extensively evaluate \emph{AutoPlan} under various deployment scenarios, in terms of both coverage and capacity.
The evaluation results show that \emph{AutoPlan} flexibly adapts to different scenarios and achieves performance comparable to exhaustive search, while requiring less than 2\% of its computation time.
\end{abstract}




\section{Introduction}
\label{sec:introduction}
Network planning constitutes a fundamental stage in the design of cellular networks, wherein critical base station (BS) parameters (e.g.,  geographic placement, antenna orientation, transmission power, and tilt) are optimized to ensure desired network coverage and performance.
With the densification of cellular networks~\cite{andreev2019future}, network planning has become increasingly intricate, requiring the joint optimization of heterogeneous base stations under site-specific propagation effects induced by irregular 3D terrain, complex building geometries, and diverse material compositions.
Conventional cellular network planning relies on a combination of field measurements (e.g., drive tests) and model-based simulations (e.g., calibrated radio propagation models), which is laborious and time-consuming. 

Digital radio twin (DRT)~\cite{khan2022digital} emerges as a promising technique to facilitate network planning.
DRT aims to replicate the real-world radio propagation in the scene, with the attributes of fidelity (imitating as closely as possible), tractability (querying as cheaply as possible), and synchronicity (tracking as timely as possible)~\cite{almasan2022network}.
By leveraging a DRT, network operators can perform trial-and-error evaluations to optimize base station parameters and assess coverage, capacity, and interference trade-offs, prior to costly field implementation.

However, it is challenging to derive a DRT with all the essential attributes.
On the one hand, existing simulators (e.g., Sionna RayTracing~\cite{hoydis2023sionna} and Wireless InSite) suffer from non-trivial simulation-to-reality (sim-to-real) discrepancy~\cite{manalastas2024simulators}, necessitating extensive calibration to improve fidelity.
On the other hand, the synchronicity attribute is difficult to achieve, as it is not cost-efficient to conduct periodic field measurements for simulator calibration.
Note that, given the complex computation (e.g., raytracing and radio effects), we can only obtain the network performance (e.g., coverage and capacity) after querying the DRT. 
Hence, it is difficult to derive the optimal base station parameters, under high-dimensional combination spaces, even if the DRT can be queried in seconds.

In this paper, we propose \emph{AutoPlan}, a new automatic network planning framework by leveraging digital radio twin techniques.
On the one hand, we derive the DRT with the following steps. 
First, we crowdsource abundant real-world data from mobile users under existing deployed base stations (e.g., prior deployments or those of other network operators). 
Second, with the online data, we continuously calibrate the parameters (particularly, the material of buildings) of the existing simulator (i.e., Sionna RT), to reduce the sim-to-real discrepancy.
Third, we accelerate the query of the calibrated simulator with GPU parallel computing to improve its tractability.
On the other hand, we design a network planning algorithm to derive the optimal base station parameters (particularly, location and power).
Specifically, we leverage Bayesian optimization to tackle the non-negligible querying costs of the DRT by creating a Gaussian process as the surrogate model and adopting expected improvement (EI) as the acquisition function.
Using the field measurement from Husker-Net (a private 5G network at the University of Nebraska-Lincoln), we extensively evaluate \emph{AutoPlan} under different deployment scenarios, in terms of both coverage, capacity, and efficiency.

Overall, we summarize the contribution of this paper as:
\begin{itemize}
    \item A digital radio twin (DRT) with the attribute of fidelity, tractability, and synchronicity; 
    \item An automatic network planning algorithm based on Bayesian optimization, that derives optimal base station parameters assisted by the derived DRT;  
    \item An extensive campus-scale evaluation, showing comparable network performance with state-of-the-art works.
\end{itemize}

\section{Digital Radio Twin}
\label{sec:twinning}
In this section, we introduce the design of the DRT by finetuning the parameters of the target scenario in the simulator.

\subsection{Problem}
The goal is to derive the DRT to achieve all the attributes of fidelity, tractability, and synchronicity. 
We consider the scene given with 3D terrain, including the full set of building objects within the region of interest. 
Each building is associated with an initial material label, such as marble or concrete, which corresponds to a specific set of material parameters. 
These material labels serve as the initial configuration for the DRT simulation. 
Since these default values may not fully reflect the actual propagation environment, a calibration process is subsequently performed to refine the material parameters using real-world signal measurements, thus improving the fidelity and reliability of the DRT.

In this work, we focus on calibrating two key electromagnetic parameters used in DRT modeling for objects within the region $\mathcal{A}$: conductivity (in S/m) and relative permittivity. 
Specifically, conductivity characterizes how a material attenuates or reflects incident electromagnetic energy. At the same time, relative permittivity affects the bending and delay of waves as they traverse the medium, both playing critical roles in DRT modeling. 
We use $\boldsymbol{\sigma} = [\sigma_1, \sigma_2, \cdots, \sigma_K ]$ and $\boldsymbol{\epsilon} = [ \epsilon_1, \epsilon_2, \cdots, \epsilon_K ]$ to denote the conductivity and relative permittivity of $K$ objects, respectively. 
The parameter set is thus defined as:
\begin{equation}
\boldsymbol{\Theta} =
\begin{bmatrix}
\sigma_1 & \sigma_2 & \cdots & \sigma_K \\
\epsilon_1 & \epsilon_2 & \cdots & \epsilon_K
\end{bmatrix}.
\end{equation}
Based on the typical electromagnetic properties of common materials~\cite{series2015effects}, we impose the following constraints on the tunable parameters: $\sigma_k \in (0, 2)$ and $\epsilon_k \in (1, 6)$.

Hence, the goal is further concretized to seek the optimal simulation parameters to minimize the sim-to-real discrepancy.

\subsection{Challenge}

Although DRT systems can simulate real-world wireless propagation, discrepancies often arise between their simulated outputs and actual measurements—referred to as the sim-to-real gap. 
This gap does not stem from a single source but rather results from the complex interplay of multiple factors, such as multipath effects, material properties, and environmental geometry. 
Due to this multifactor coupling, the gap cannot be eliminated by simply adjusting a single parameter. 
The key challenge is developing a systematic methodology for characterizing and mitigating this complex, coupled discrepancy.

\subsection{Solution}
To address the above problem, we develop a material learning method by leveraging the differentiable property of state-of-the-art simulators.

To calibrate $\boldsymbol{\Theta}$, we utilize both the measured and simulated reference signal received power (RSRP, in dBm) values. Let $\hat{\mathbf{r}} = [\hat{r}_1, \hat{r}_2, \cdots, \hat{r}_P]$ represent the RSRP collected at $P$ locations in $\mathcal{A}$ from existing BSs. The corresponding simulated RSRP vector, computed using the DRT under $\boldsymbol{\Theta}$, is denoted by $\tilde{\mathbf{r}}(\boldsymbol{\Theta}) = [\tilde r_1(\boldsymbol{\Theta}), \tilde r_2(\boldsymbol{\Theta}), \cdots, \tilde r_P(\boldsymbol{\Theta})]$. 
The calibration task is defined as
\begin{equation}
\label{eq2}
\boldsymbol{\Theta}^* = \arg\min_{\boldsymbol{\Theta}} \| \tilde{\mathbf{r}}(\boldsymbol{\Theta}) - \hat{\mathbf{r}} \|,
\end{equation}
where $\| \cdot \|$ represents the $\ell_2$ norm between the simulated and measured RSRP vectors. In practice, we minimize the empirical loss: 
\begin{equation}
    \mathcal{L}(\boldsymbol{\Theta}) = \frac{1}{P} \sum\nolimits_{p=1}^P \left\| \tilde{\mathbf{r}}(\boldsymbol{\Theta}) - \hat{\mathbf{r}} \right\|.
\end{equation}
We optimize this loss over $E$ training epochs using stochastic gradient descent~\cite{boyd2010convex}. At each epoch $e \in \{1, \dots, E\}$, the parameters are updated as:
\begin{equation}
\label{eq: eq8}
\boldsymbol{\Theta}^{(e)} \leftarrow \boldsymbol{\Theta}^{(e-1)} - \eta \nabla \mathcal{L}(\boldsymbol{\Theta}^{(e-1)}) ,
\end{equation}
where $\eta$ is the learning rate and $\boldsymbol{\Theta}^{(e)}$ denotes the parameters after the $e$-th epoch. In practice, after $E$ epochs of SGD, we use the checkpoint that yields the lowest training loss as our estimate of the optimizer in (\ref{eq2}), denoted by $\boldsymbol{\Theta}^*$.

\section{Automatic Network Planning}
\label{sec:planning}
In this section, we introduce the automatic network planning algorithm by leveraging the aforementioned DRT.

\subsection{Problem}
The goal of network planning is to obtain the optimal base station parameters (e.g., location and orientation) that maximize the specified network performance (e.g., coverage and capacity) in the aforementioned DRT, under a given geographic area.

\begin{figure}[!t]
	\centering
	\includegraphics[width=3.0in]{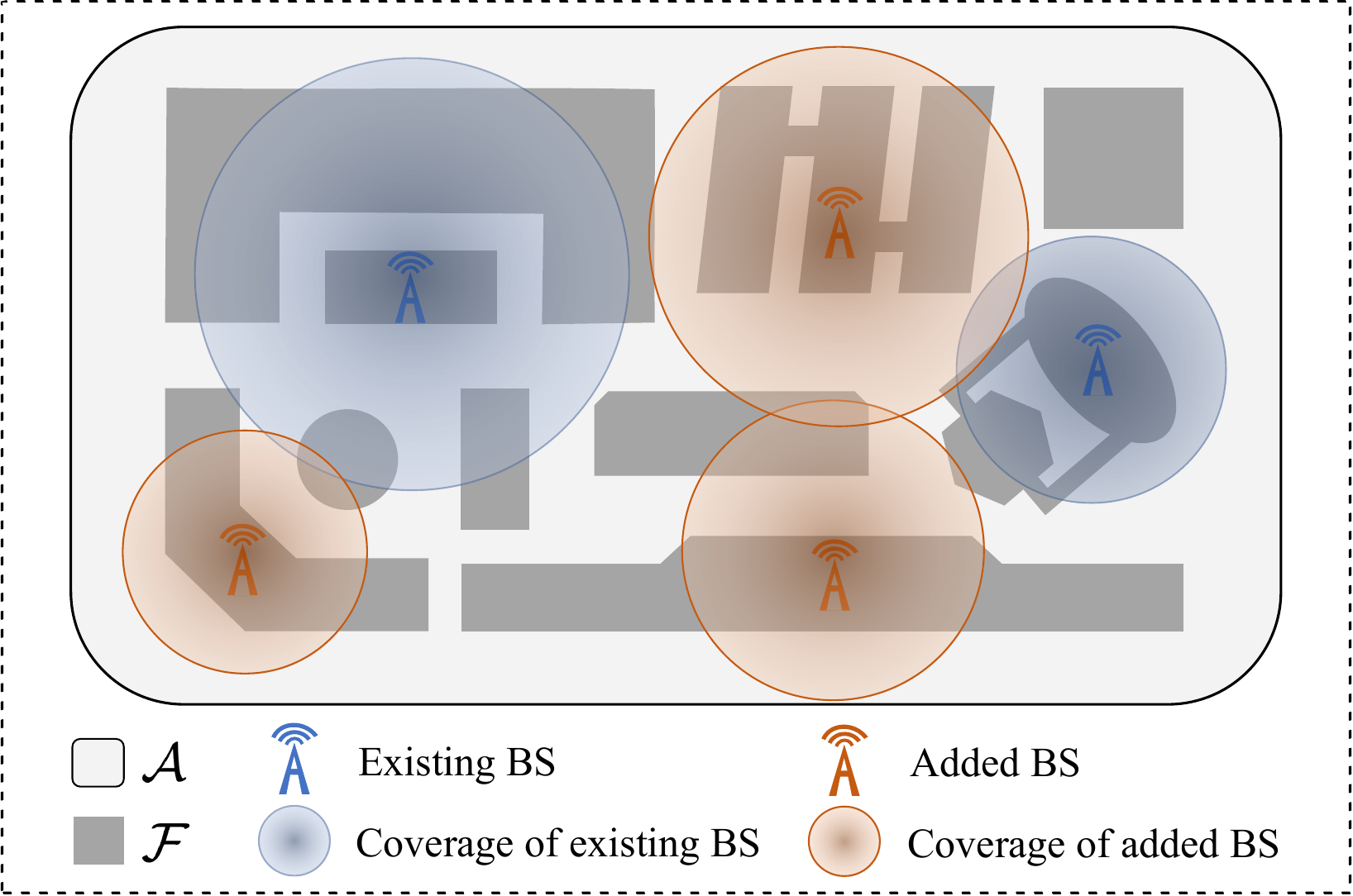}
	\caption{Illustration of BSs deployment in feasible region $\mathcal{F}$.}
	\label{fig:figure1}
\end{figure}

As illustrated in Fig. \ref{fig:figure1}, we consider a geographic region $\mathcal{A} \subset \mathbb{R}^2$ with given 3D terrain, such as buildings, trees, and walkways.
Here, we denote $\hat{\mathcal{B}} = {(\hat{x}_m, \hat{y}_m)}_{m=1}^M$ as the deployed $M$ base stations (BSs).  
To further improve coverage and expand capacity, we aim to deploy additional $N$ BSs at locations $\bar{\mathcal{B}} = {(\bar{x}_n, \bar{y}_n)}_{n=1}^N$, selected from a feasible region $\mathcal{F} \subseteq \mathcal{A}$ that meets practical placement constraints (e.g., rooftop availability and height limits). 
The complete BS set after deployment is $\mathcal{B} = \hat{\mathcal{B}} \cup \bar{\mathcal{B}}$, and the goal is to determine $\bar{\mathcal{B}}$ to maximize the overall performance of the network in $\mathcal{A}$.

Next, we introduce two network performance metrics, including coverage and capacity. 
We discretize $\mathcal{A}$ into a uniform square grid with resolution $a\times a$ meters, yielding
$L = \left\lceil |\mathcal{A}|/{a^2} \right\rceil$ sampling points, where $|\mathcal{A}|$ is the area of $\mathcal{A}$, and $\lceil\cdot\rceil$ represents the ceiling function (i.e., rounding up to the nearest integer).
For each sampling point $\ell \in \{1, \dots, L\}$, the characteristics of the received signal are obtained using the aforementioned DRT that reflects the real-world propagation environment. 

We evaluate network coverage using $r_\ell(\boldsymbol{\Theta})$, which denotes the received power of the reference signal (RSRP) at the sampling point $\ell$, corresponding to the strongest signal received from all BSs in the set $\mathcal{B}$.
In addition, we evaluate the network capacity using the signal-to-noise ratio (SNR), which captures signal quality in the presence of interference. 
Let $s_\ell(\boldsymbol{\Theta})$ represent the SNR at point $\ell$, computed under the current parameter setting $\boldsymbol{\Theta}$ in the DRT.
Given the parameter of the DRT, we can obtain RSRP and SNR values at all sampling points, and then define the following performance metrics. 
Given a predefined RSRP threshold $r_{\mathrm{th}}$, the coverage rate is defined as:
\begin{equation}
C(\mathcal{B}) = \frac{1}{L} \sum\nolimits_{\ell=1}^{L} \mathds{1}\left[ r_\ell(\boldsymbol{\Theta}) > r_{\mathrm{th}} \right],
\end{equation}
where $\mathds{1}[\cdot]$ is the indicator function, returning 1 if the condition is satisfied and 0 otherwise. 
The network capacity, reflecting the average spectral efficiency across all sampling points, is expressed by:
\begin{equation}
S(\mathcal{B}) = \frac{1}{L} \sum\nolimits_{\ell=1}^{L} \log_2\left(1 + s_\ell(\boldsymbol{\Theta})\right).
\end{equation}
Finally, we combine these two metrics into a composite objective function:
\begin{equation}
\label{eq:eq3}
T(\mathcal{B}) = \beta \cdot C(\mathcal{B}) + S(\mathcal{B}),
\end{equation}
where $\beta > 0$ is a tunable weight controlling the trade-off between coverage and capacity. 
Here, a larger $\beta$ places greater emphasis on coverage, whereas a smaller one prioritizes capacity.

Hence, the goal of network planning is further concretized to determine the optimal set of additional deployment locations $\bar{\mathcal{B}}^*$ that maximizes the overall objective:
\begin{equation}
\label{eq:eq4}
\bar{\mathcal{B}}^* = \arg\max_{\bar{\mathcal{B}} \subseteq \mathcal{F}} T(\hat{\mathcal{B}} \cup \bar{\mathcal{B}}),
\end{equation}
where $\hat{\mathcal{B}}$ denotes the set of existing BSs and $\mathcal{F} \subseteq \mathcal{A}$ represents the feasible deployment region.

\subsection{Challenge}
The key technical challenge lies in the unknown correlation between the input of base station locations and the output of the network performance.
This is attributed to the complex computation (e.g., ray tracing and radio effects) in the DRT, which cannot be mathematically represented.
Specifically, we can obtain only the output of network performances by querying the DRT (non-negligible querying costs) under the input base station locations.
In other words, the objective function $T(\mathcal{B})$ is unknown, which falls into the realm of blackbox optimization.

\subsection{Solution}
To address the above problem, we design an automatic network planning method by leveraging Bayesian optimization (BO).
BO~\cite{wang2023recent, garnett2023bayesian} is the state-of-the-art global searching approach (especially for handling expensive querying costs), generally consisting of a surrogate model and acquisition function.
The surrogate model aims to online learn the black box function (e.g., mean and variance), where candidates include Gaussian process~\cite{rasmussen2003gaussian} and Bayesian neural networks~\cite{li2023study}.
Based on the updated surrogate model, the acquisition function evaluates the customized utility of different actions (i.e., BS locations) in the action space (i.e., $\mathcal{F}$).
In each iteration, the next action is selected by maximizing the acquisition function (e.g., expected improvement (EI) or probability of improvement (PI))~\cite{frazier2018tutorial}.

To support multiple base station deployment, a naive approach in BO is to use the surrogate model to learn the combinational action space (i.e., all candidate BS locations) and overall network performance.
However, this naive approach exponentially expands the action space and practically leads to a much slower convergence speed and more cumulative querying costs, in terms of time consumption and computation complexity.
Hence, we convert the problem to a series of incremental base station deployment problems.
In other words, we apply Bayesian optimization only to search for one additional base station deployment at a time.
Once the base station location is determined, we put it into the existing deployed base station sets (i.e., $\hat{\mathcal{B}}$), and then create another Bayesian optimization searching for the next base station.
This incremental deployment continues until all $N$ base stations are deployed.

In BO, the true objective function \eqref{eq:eq3} is replaced by a surrogate model that is cheaper to evaluate but still captures the underlying spatial correlation of the objective.
Specifically, we use a Gaussian process (GP) as the surrogate model, which is widely adopted and time-evaluated in multiple application domains.
Here, we model the target function as:
\begin{equation}
T(\hat{\mathcal{B}} \cup \bar{\mathcal{B}}) \sim \mathcal{GP}\left(\mu(\bar{\mathcal{B}}), \sigma^2(\bar{\mathcal{B}}) \right),
\end{equation}
where $\mu(\bar{\mathcal{B}})$ and $\sigma^2(\bar{\mathcal{B}})$ are the posterior mean and variance of the GP given the observations collected in previous iterations.
The GP surrogate allows us to predict both the expected performance of a candidate deployment $\bar{\mathcal{B}}$ and the uncertainty of this prediction, which is crucial for balancing exploration and exploitation.

The next action to query is generally determined by maximizing the selected acquisition function, which defines the utility of all actions in the action space while balancing exploration and exploitation. 
Specifically, we adopt the EI as the acquisition function, which quantifies the expected gain over the current best performance $T_{\text{best}}$:
\begin{equation}
\alpha(\bar{\mathcal{B}}) = \mathbb{E} \left[ \max\left(0, T(\hat{\mathcal{B}} \cup \bar{\mathcal{B}}) - T_{\text{best}}\right) \right].
\end{equation}
This criterion favors actions that are likely to improve the objective while still exploring regions of high uncertainty.
At each iteration $t$, we select the next set of deployment locations by maximizing the acquisition function over the feasible region $\mathcal{F}$:
\begin{equation}
\bar{\mathcal{B}}_{t+1} = \arg\max{\bar{\mathcal{B}} \subseteq \mathcal{F}} \alpha (\bar{\mathcal{B}}).
\end{equation}
The newly selected action is then evaluated using the derived DRT model (see Sec.~\ref{sec:twinning}), and the resulting observation is incorporated into the GP to update its posterior distribution. 
This iterative process continues until a stopping criterion, such as a fixed iteration limit or convergence threshold, is met.

\section{The AutoPlan Algorithm}
\label{sec:solution}


In this section, we summarize the workflow of the \emph{AutoPlan} algorithm in Alg.~\ref{alg:algorithm1}.
Overall, the algorithm takes the input of 3D terrain of the geographic region and the location of deployed base stations $\hat{\mathcal{B}}$, and generates the output of the location of the $N$ additional base stations. 
In the digital radio twinning stage, it observes real-world RSRPs $\hat{\mathbf{r}}$ from crowdsourced mobile users and updates the building materials.
Specifically, it iteratively calculates the loss function by comparing the real-world RSRPs with the simulated RSRPs from the simulator, and performs multiple gradient descent steps.
In the stage of network planning, it iteratively searches the location of the next base station by 1) selecting a base station location (i.e., maximizing the EI acquisition function); 2) querying the DRT to obtain the network performance; and 3) updating the GP surrogate model with new observations.

\begin{algorithm}[!h]
\caption{The \emph{AutoPlan} Algorithm}
\label{alg:algorithm1}
    
    \KwIn{
    Region $\mathcal{A}$, feasible subregion $\mathcal{F}$, \\
    \quad Existing BS set $\hat{\mathcal{B}} = \{(\hat{x}_m, \hat{y}_m)\}_{m=1}^M$,\\
    \quad Measured RSRP $\hat{\mathbf{r}}$, simulated RSRP $\tilde{\mathbf{r}}(\boldsymbol{\Theta})$,\\
    \quad Grid size $a$, RSRP threshold $r_{\text{th}}$,\\
    \quad Number of new BSs $N$, training epochs $E$
    }
    
    \KwOut{
    Calibrated parameters $\boldsymbol{\Theta}^*$, new BS locations $\bar{\mathcal{B}} = \{(\bar{x}_n, \bar{y}_n)\}_{n=1}^N$
    }
    
    \textbf{Stage 1: Digital radio twin calibrating}\;
    Initialize $\boldsymbol{\Theta}$ randomly\;
    \For{$t \gets 1$ \KwTo $E$}{
        Compute simulated RSRP $\tilde{\mathbf{r}}(\boldsymbol{\Theta})$ using DRT\;
        Compute loss $\mathcal{L}(\boldsymbol{\Theta})$ between $\hat{\mathbf{r}}$ and $\tilde{\mathbf{r}}(\boldsymbol{\Theta})$\;
        Update parameters $\boldsymbol{\Theta}$\ according to (\ref{eq: eq8});
    }
    Keep the checkpoint with the minimum training loss over $E$ epochs, and denote it as $\boldsymbol{\Theta}^*$;
    
    \vspace{1mm}
    \textbf{Stage 2: Automatic network planning}\;
    Initialize $\bar{\mathcal{B}} \gets \emptyset$, observation set $\mathcal{D} \gets \emptyset$\;
    
    \For{$n \gets 1$ \KwTo $N$}{
        Fit GP surrogate model to $\mathcal{D}$\;
        Compute acquisition function $\alpha(\bar{b})$ using Expected Improvement ($\bar{b}=(\bar x,\bar y)$)\;
        Select next BS: $\bar{b}^* \gets \arg\max\limits_{\bar{b} \in \mathcal{F} \setminus (\hat{\mathcal{B}} \cup \bar{\mathcal{B}})} \alpha(\bar{b})$\;
        Evaluate $T(\hat{\mathcal{B}} \cup \bar{\mathcal{B}} \cup \{\bar{b}^*\})$ using calibrated DRT\;
        Update: $\bar{\mathcal{B}} \gets \bar{\mathcal{B}} \cup \{\bar{b}^*\}$\;
        Add $(\bar{b}^*, T)$ to $\mathcal{D}$\;
    }

\Return{$\boldsymbol{\Theta}^*$, $\bar{\mathcal{B}}$}
\end{algorithm}

\section{Performance Evaluation}
In this section, we conduct extensive simulations to evaluate \emph{AutoPlan} by comparing with state-of-the-art works.

\textbf{Parameters.}
We select a $1210~\text{m}\times 1138~\text{m}$ campus area at the University of Nebraska–Lincoln, i.e., the deployment region $\mathcal{A}$. There are $K=127$ objects (i.e., buildings) in this region.
We obtain the scene geometry from OpenStreetMap3D and derive the feasible deployment region $\mathcal{F}\subseteq\mathcal{A}$ by extracting the union of building footprints. 
One existing private 5G base station is located near the southwest corner of $\mathcal{A}$, yielding the initial BS set $\hat{\mathcal{B}}={(\hat{x}_m,\hat{y}_m)}_{m=1}^{M}$ with $M=1$.
In the region $\mathcal{A}$, we collect $P = 1494$ RSRP measurements $\hat{\mathbf r}=[\hat r_1,\hat r_2,\ldots,\hat r_P]$ via drive testing, with a 5G dongle (Quectel RM520N-GL).
The accurate location of the receiver is obtained by using RTK-GNSS, providing horizontal accuracy better than $2$ cm. 
The deployed base stations are equipped with a planar antenna array configured as a $1 \times 2$ horizontal linear array with half-wavelength element spacing. 
The array adopts the 3GPP TR 38.901 directional antenna pattern and vertical polarization, representing a typical 5G small cell configuration. 
In the evaluation, we aim to select $N=5$ additional BS locations from the feasible region, i.e., $\bar{\mathcal{B}}={(\bar{x}_n,\bar{y}_n)}_{n=1}^{5}\subseteq\mathcal{F}$. 
To calculate coverage, we set $r_{th} = -90 \mathrm{dBm}$ and $a = 0.2 \mathrm{m}$.
The default maximum transmission power of the base station is $43\,\mathrm{dBm}$. Additionally, we set $\beta = 10$ to balance the contributions of coverage and capacity in the target objective.
All simulations were conducted on a desktop with an Intel Core i7 CPU, an NVIDIA RTX 3080 GPU, and Ubuntu 24.04 OS, using Sionna RT v1.1.0~\cite{hoydis2023sionna}. 
Specifically, we calculate the RSRP and SNR by using \emph{RadioMapSolver} on a uniform $0.2\mathrm{m}\times0.2\mathrm{m}$ grid.
To learn the building materials, we use the Adam optimizer~\cite{kingma2014adam} with learning rate $\eta=0.01$, for $E=300$ epochs. 


We compare \emph{AutoPlan} with the following methods:
\begin{itemize}
\item \textbf{\emph{Random Sampling (RS).}} At each selection step, we randomly sample feasible candidate locations from $\mathcal{F}$ and evaluate $T(\cdot)$ using the calibrated DRT. For a fair comparison with \emph{AutoPlan}, we use the same pool of 100 candidate BS groups and report the group with the largest objective as the final solution.
\item \textbf{\emph{Exhaustive Search (ES).}} ES discretizes the feasible action space $\mathcal{F}$ using a $5~\mathrm{m}$ step factor, evaluates the objective $T(\cdot)$ at every candidate point under the calibrated DRT, and selects the location with the highest performance. 
ES provides a near-optimal result under the given resolution with the highest computational complexity.
\end{itemize}

\subsection{Digital Radio Twinning}
\begin{figure*}[!t]
  \captionsetup{justification=centering}
  \centering
  \begin{minipage}[t]{0.24\textwidth}
    \centering
    \includegraphics[width=\linewidth]{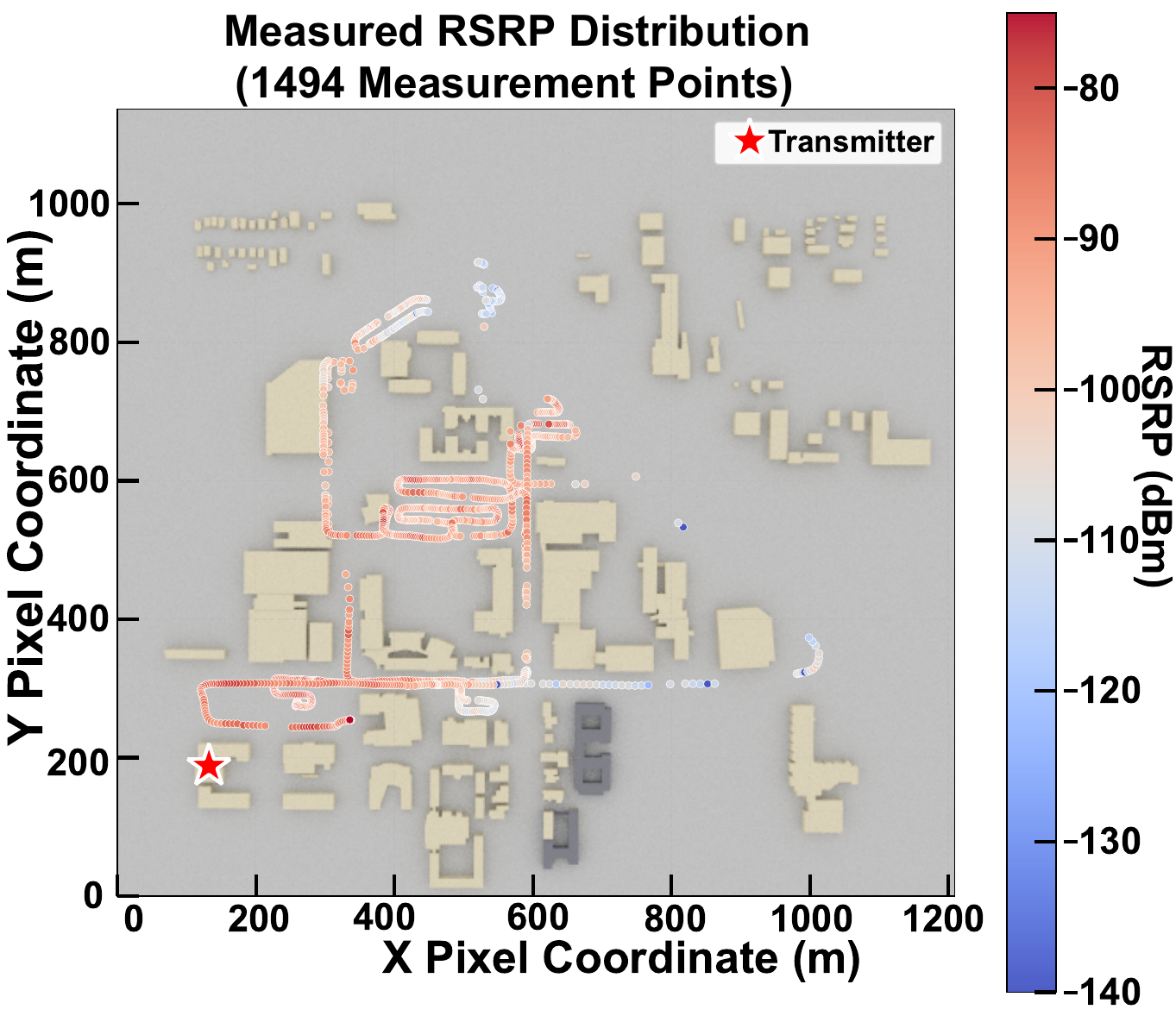}
    \caption{The trajectory of measured RSRP.}
    \label{fig:figure2}
  \end{minipage}
  \hfill
  \begin{minipage}[t]{0.24\textwidth}
    \centering
    \includegraphics[width=\linewidth]{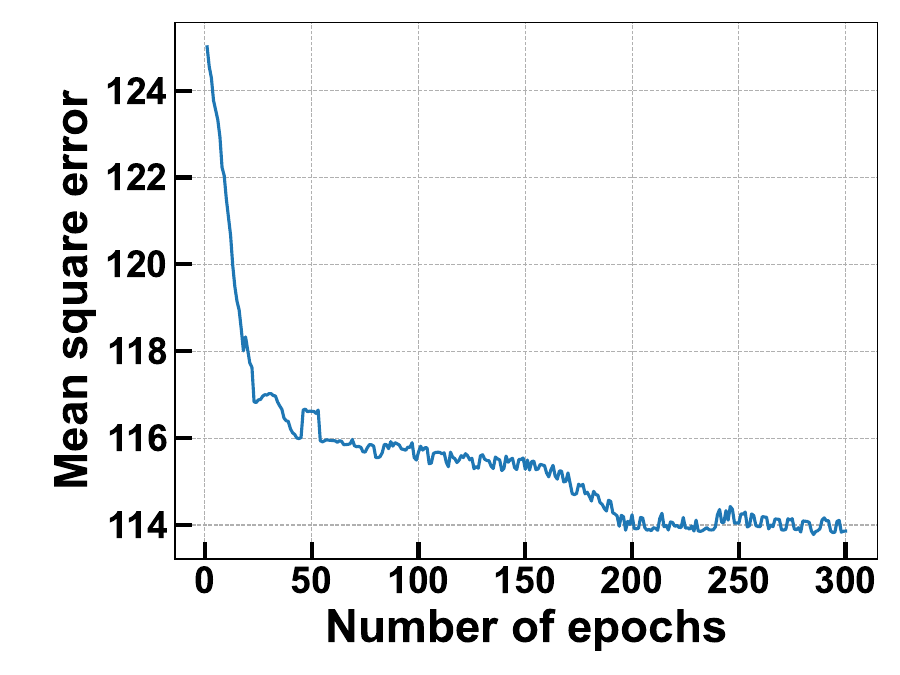}
    \caption{The MSE loss during DRT calibration.}
    \label{fig:fig3_1}
  \end{minipage}
  \hfill
  \begin{minipage}[t]{0.24\textwidth}
    \centering
    \includegraphics[width=\linewidth]{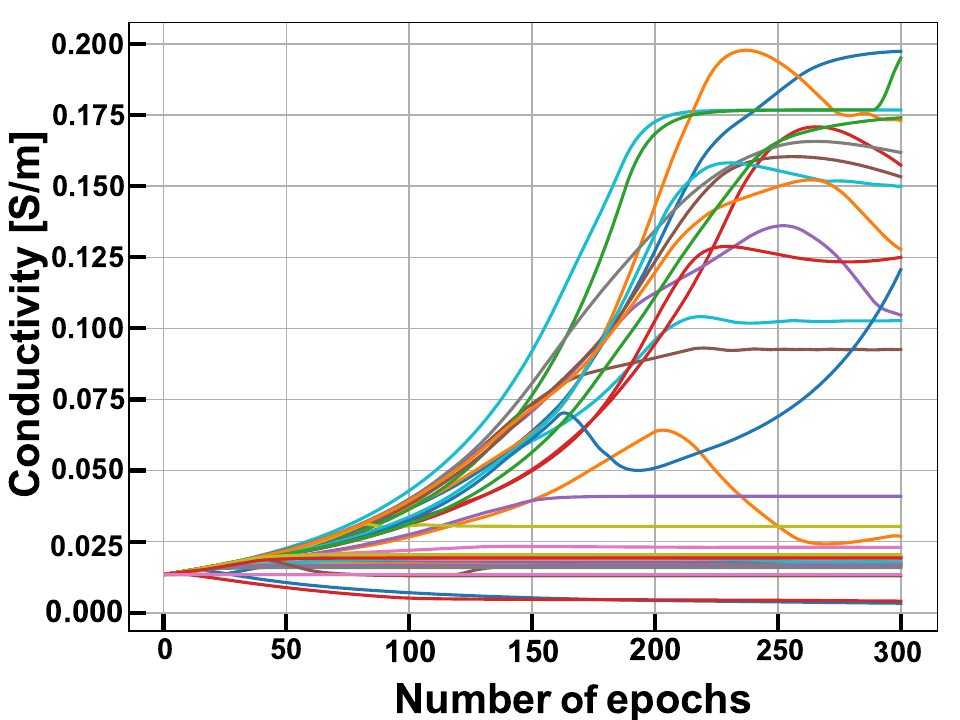}
    \caption{Conductivity of building materials, during DRT calibration.}
    \label{fig:fig3_2}
  \end{minipage}
  \hfill
  \begin{minipage}[t]{0.24\textwidth}
    \centering
    \includegraphics[width=\linewidth]{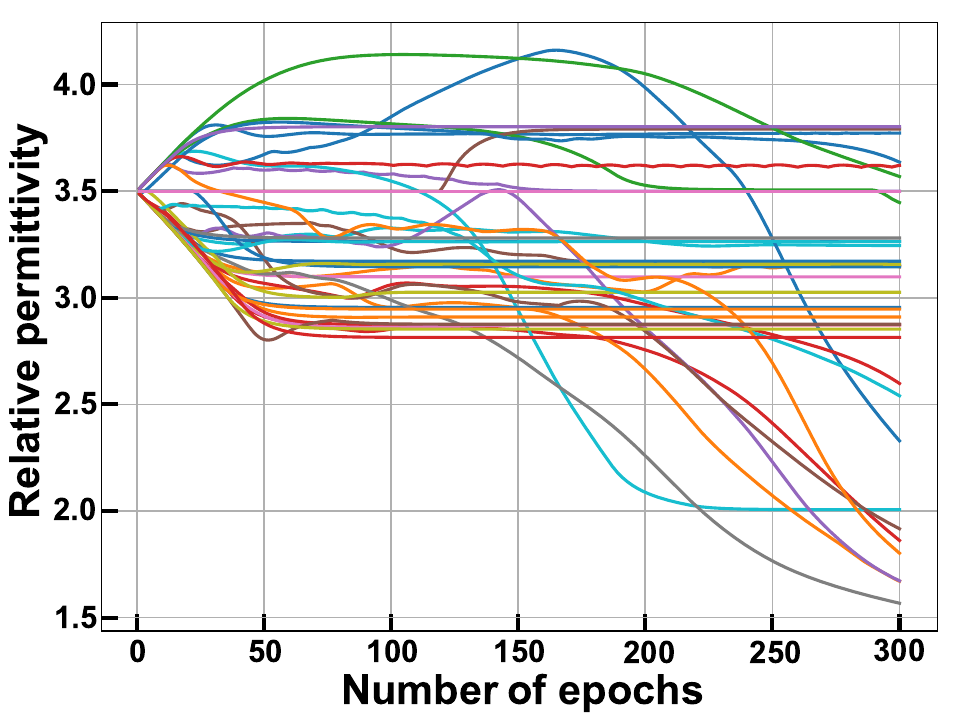}
    \caption{Relative permittivity of building materials, during DRT calibration.}
    \label{fig:fig3_3}
  \end{minipage}
\end{figure*}

In this subsection, we evaluate the DRT for all the attributes of fidelity, tractability, and synchronicity, under the \emph{AutoPlan} algorithm.
Specifically, we collect the field measurement via a drive test in the City Campus of the University of Nebraska-Lincoln, as shown in Fig.~\ref{fig:figure2}. Fig. \ref{fig:fig3_2} and \ref{fig:fig3_3} show the learned conductivity and permittivity during DRT calibration.



As shown in Fig. 3, the training MSE loss steadily decreases over the 300 gradient descent steps and converges after around 200 steps, with a minimum value of $113.78$. In other words, the sim-to-real discrepancy is reduced by $11.23 (8.98\%)$, by calibrating the parameters of building materials.

Note that it is impossible to reduce the MSE loss to zero, because the sim-to-real discrepancy is attributed to multiple complex factors, including but not limited to building materials, simulator abbreviation, radio propagation models, etc., losses, circuit noises, and much more. In our hardware setups, the time consumption of a single gradient descent step is $4.97 \mathrm{s}$ on average, with the variance of $0.07$. This indicates that, with sufficient GPU resources, the DRT can be continuously calibrated and updated in real time.

\vspace{0.5 em}
\begin{table}[!h]
\renewcommand\arraystretch{1.0}
    \caption{DRT inference time.} \centering
    \label{tab:TableI} 
	\setlength{\tabcolsep}{2.5mm}{
	\begin{tabular}{cccccc}
		\toprule
		Number of BSs & 1 & 2 & 3 & 4 & 5 \\ 
		\midrule
		Avg. time ($\mathrm{s}$) & 0.39 & 0.72 & 0.87 & 1.07 & 1.24 \\
 
		\bottomrule
	\end{tabular}}
\end{table}
\vspace{0.5 em}

In addition, Table~\ref{tab:TableI} reports the average inference time of the DRT for varying numbers of deployed base stations. Notably, it is $0.39\mathrm{s}$ with a single base station and increases to $1.24,\mathrm{s}$ with five, highlighting the computational cost of large-scale evaluations.

\subsection{Automatic Network Planning}
In this subsection, we evaluate the network performance of network planning under all comparison solutions. 


\begin{figure}[!t]
  \captionsetup{justification=centering}
  \centering
  \begin{minipage}[t]{0.24\textwidth}
    \centering
    \includegraphics[width=\linewidth]{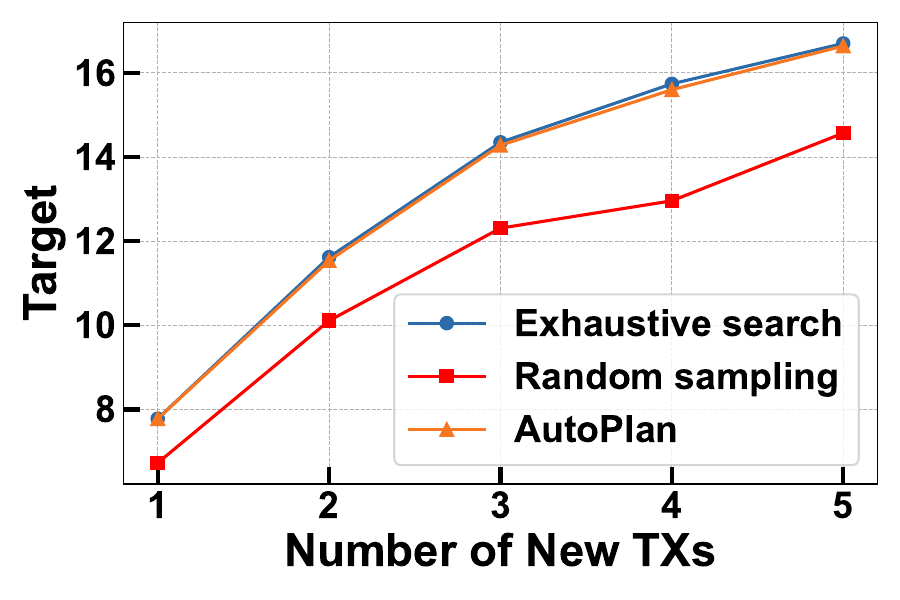}
    \caption{Network performance under different methods.}
    \label{fig:figure4}
  \end{minipage}\hfill
  \begin{minipage}[t]{0.24\textwidth}
    \centering
    \includegraphics[width=\linewidth]{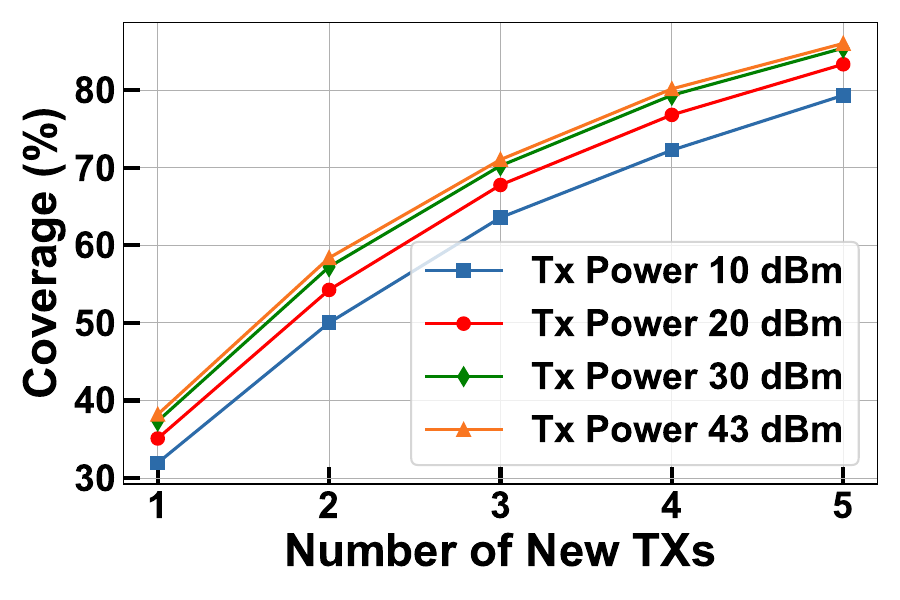}
    \caption{Network performance under different TX powers.}
    \label{fig:figure5}
  \end{minipage}
\end{figure}

The network performance, as defined in Eq. (\ref{eq:eq3}), for all compared network planning methods is illustrated in Fig. \ref{fig:figure4}. 
As we can see, exhaustive search yields the highest overall performance, while random sampling performs significantly worse than both \emph{AutoPlan} and ES. 
Notably, \emph{AutoPlan} achieves performance nearly identical to that of ES, demonstrating its effectiveness in accurate and efficient BS deployment. In terms of computation time, ES requires $602$ minutes, approximately $58.33$ times longer than \emph{AutoPlan}, highlighting the latter’s practical efficiency.

\vspace{0.7 em}
\begin{table}[!h]
\renewcommand\arraystretch{1.0}
    \caption{Performance of deployment methods.} \centering
    \label{tab:TableII} 
	\setlength{\tabcolsep}{2.5mm}{
	\begin{tabular}{cccc}
		\toprule
		{ }& AutoPlan & ES & Random sampling \\ 
		\midrule
		Coverage ($\%$) & \textbf{86.04} & 83.78 & 80.33 \\
		Capacity (bit/s/Hz) & 8.04 & \textbf{8.32} & 6.54\\
        Target & 16.64 & \textbf{16.70} & 14.57\\ 
		\bottomrule
	\end{tabular}}
\end{table}
\vspace{0.7 em}

In Table~\ref{tab:TableII}, we further show the detailed network coverage and capacity under 5 additional base station deployments. 
We can see that, \emph{AutoPlan} achieves 99.64\% overall target performance of ES, including a slightly higher network coverage ($2.40\%$), but a lower network capacity ($3.36\%$). These results demonstrate the effectiveness of \emph{AutoPlan} in selecting near-optimal BS deployment with significantly reduced computational cost.


\begin{figure*}[!t]
  \captionsetup{justification=centering}
  \centering
  \begin{minipage}[t]{0.24\textwidth}
    \centering
    \includegraphics[width=\linewidth]{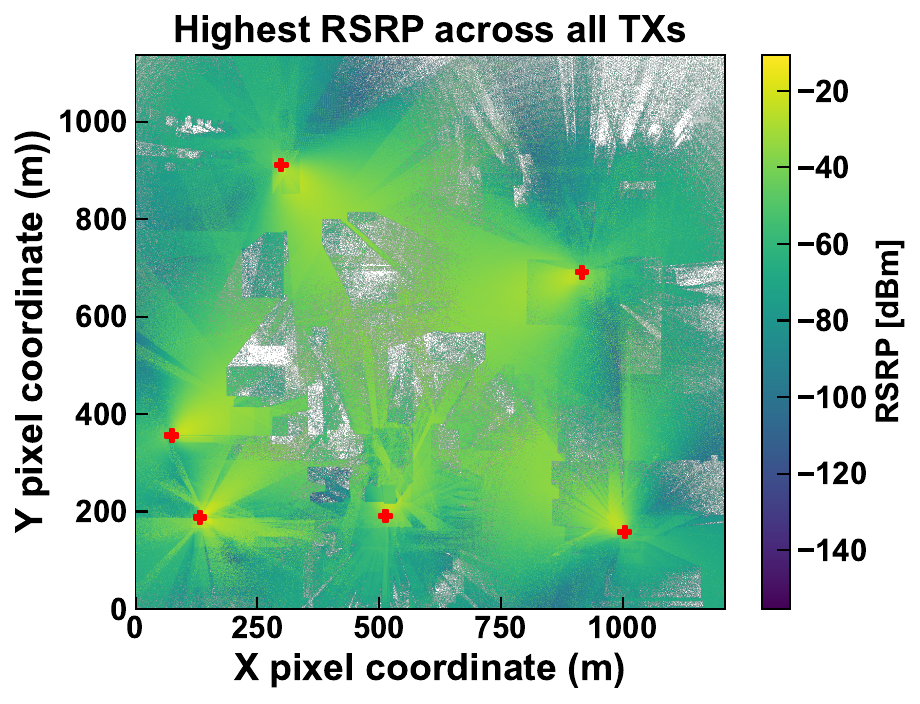}
    \caption{RSRP heatmaps of \emph{AutoPlan}.}
    \label{fig:fig7_1}
  \end{minipage}
  \hfill
  \begin{minipage}[t]{0.24\textwidth}
    \centering
    \includegraphics[width=\linewidth]{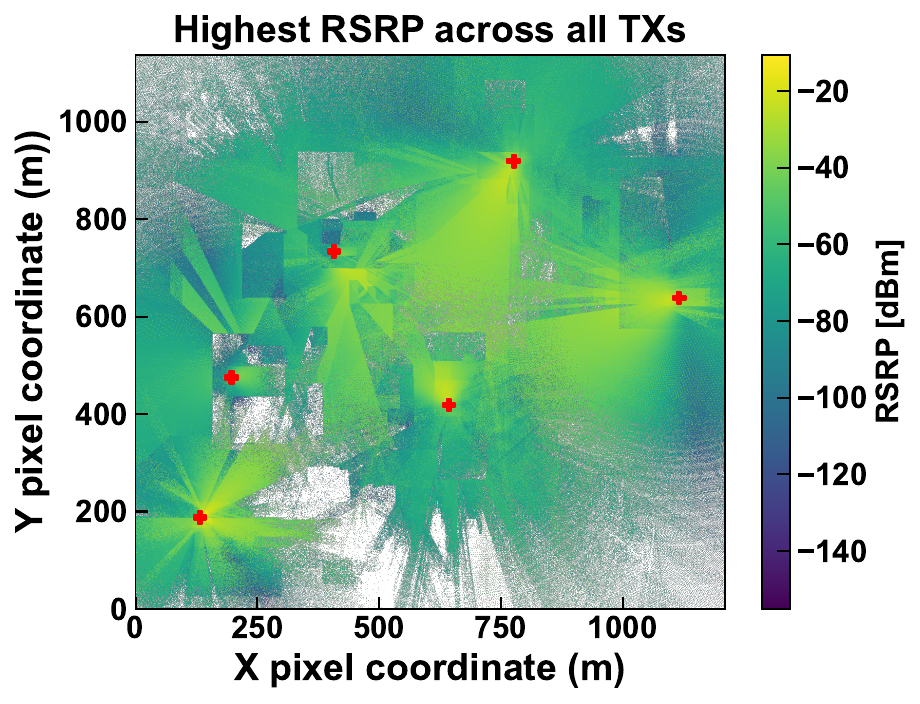}
    \caption{RSRP heatmaps of Random Sampling.}
    \label{fig:fig7_2}
  \end{minipage}
  \hfill
  \begin{minipage}[t]{0.24\textwidth}
    \centering
    \includegraphics[width=\linewidth]{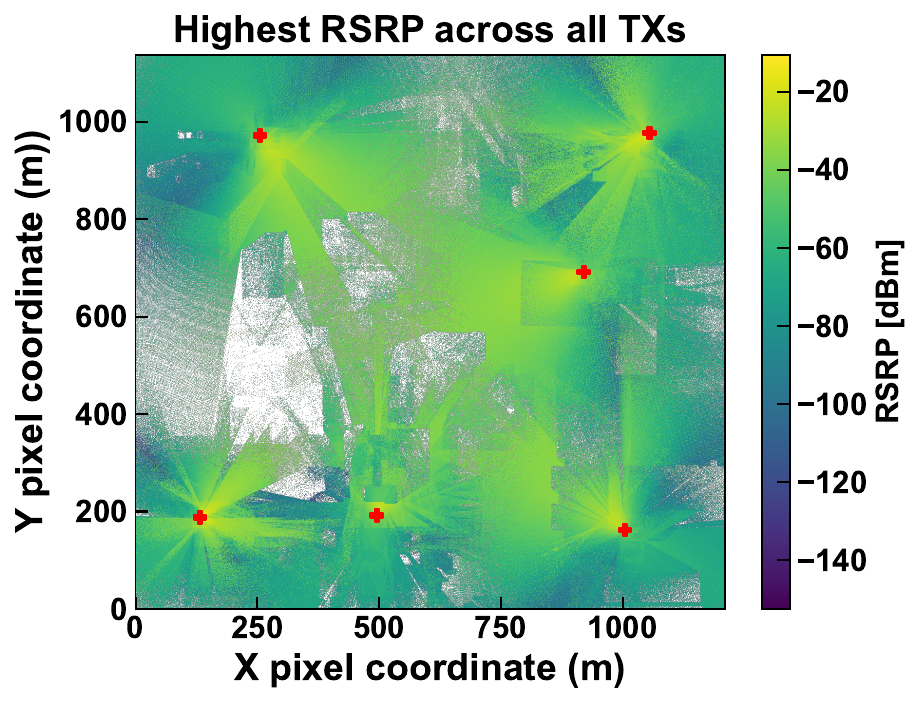}
    \caption{RSRP heatmaps of Extensive Search.}
    \label{fig:fig7_3}
  \end{minipage}
  \hfill
  \begin{minipage}[t]{0.24\textwidth}
    \centering
    \includegraphics[width=\linewidth]{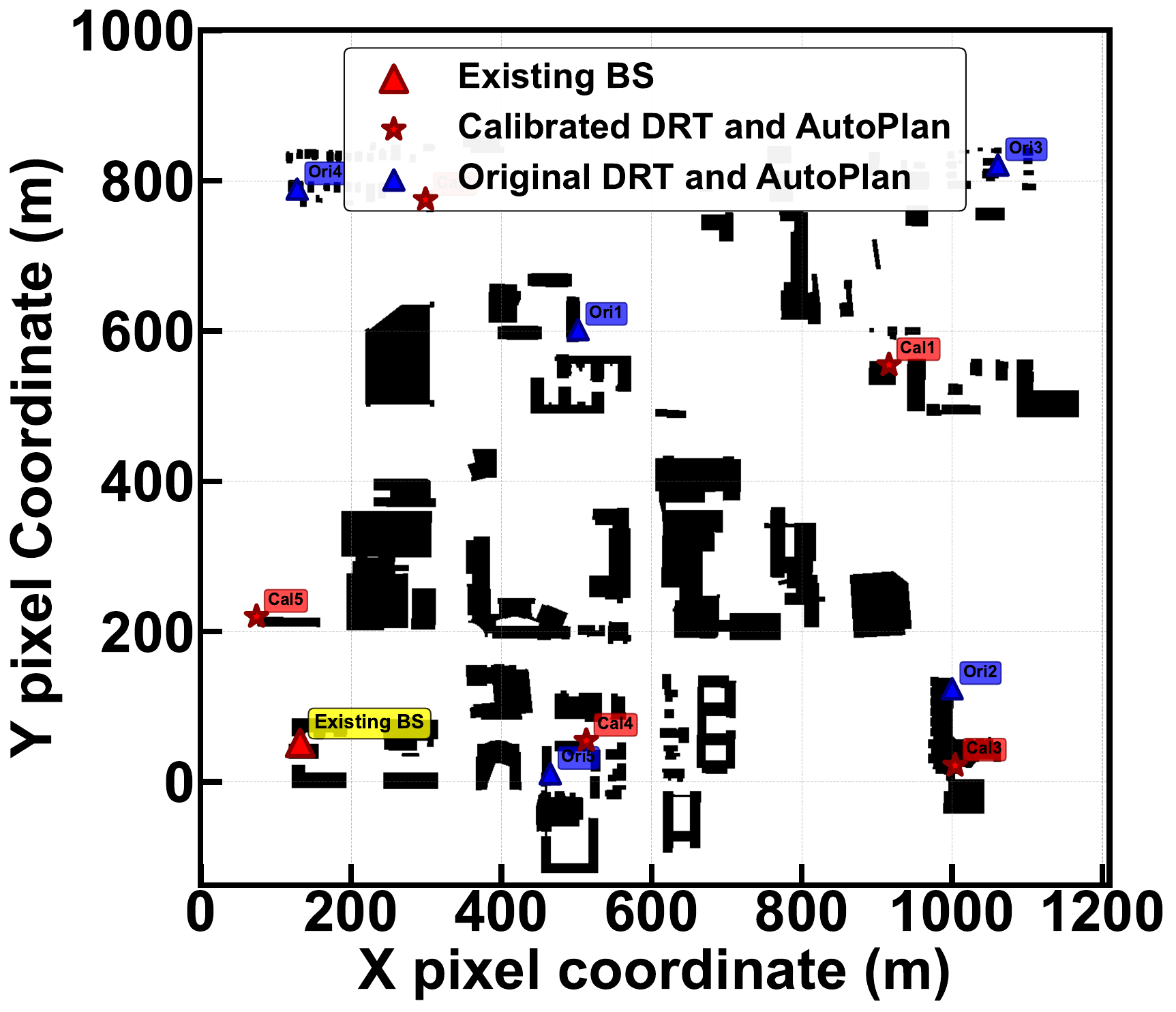}
    \caption{Base station locations of \emph{AutoPlan}.}
    \label{fig:figure6}
  \end{minipage}
\end{figure*}




In Fig.~\ref{fig:fig7_1}-\ref{fig:fig7_3}, we compare the RSRP heatmaps generated by the final BS deployments. 
Among the three methods, \emph{AutoPlan} achieves the largest coverage area with strong RSRP, while Random Sampling suffers from fragmented coverage and extensive low-signal regions. 
Moreover, four of the BSs selected by \emph{AutoPlan} are close to the placements in the ES solution, indicating that it can approximate the near-optimal deployment with far fewer evaluations.

In Fig.~\ref{fig:figure6}, we show the impact of DRT on the selection of base station location under \emph{AutoPlan}. As observed, the deployment results obtained via the original DRT differ significantly from those based on the calibrated DRT, highlighting the influence of DRT accuracy on BS placement decisions. Quantitatively, the optimal deployment selected using \emph{AutoPlan} under the original DRT achieves a target value of $18.20$, while the same method applied on the calibrated DRT yields a target of $16.64$—indicating a gap of approximately $10\%$. This result underscores the necessity of calibrating the DRT to ensure reliable and accurate network planning decisions.

Moreover, we evaluate the sensitivity of coverage to transmit power in Fig.~\ref{fig:figure5}, which plots the coverage achieved by \emph{AutoPlan} as a function of the number of newly deployed BSs under different transmit power levels. As expected, coverage improves with increased power; however, the gain becomes marginal beyond $30 \mathrm{dBm}$. For example, when five new BSs are installed, increasing the power from $30\mathrm{dBm}$ to $43\mathrm{dBm}$ only results in an improvement of $2\%$. Considering the trade-off between performance and power consumption, we adopt $43\mathrm{dBm}$ as the default setting in subsequent experiments. This value yields the highest observed coverage and corresponds to typical configurations of micro-base stations.








\section{Related Work}

Traditional approaches to BS deployment include both manual measurement and optimization-based methods. Early efforts often relied on drive testing or site surveys to evaluate signal quality at predefined locations~\cite{molisch2022wireless}, which are labor-intensive, time-consuming, and inflexible in adapting to changing user demands. To overcome these limitations, optimisation-based methods were introduced to automate the process using mathematical formulations or heuristics. 
Chiaraviglio et al.~\cite{chiaraviglio20215g} formulated BS planning as a mixed-integer linear program under service and EMF constraints, and proposed a heuristic solution named PLATEA. While effective in balancing coverage and EMF compliance, the method relies on static environmental models and sensitive parameters, limiting its adaptability in dynamic scenarios. Similarly, Philip et al.\cite{philip2023cellular} applied meta-heuristic algorithms such as particle swarm optimization (PSO) and genetic algorithms (GA) for BS placement. Although PSO achieved better coverage and efficiency than GA, both approaches depend on simplified propagation models and static user distributions, making them less suitable for realistic and data-driven deployments.

In contrast, AI/ML-based methods aim to enhance scalability and adaptability by replacing manual engineering and handcrafted models with data-driven learning. 
AutoBS~\cite{lee2025autobs} employs reinforcement learning to select BS locations, enabling fast inference during deployment. However, it requires large-scale pretraining on labled environment performance pairs, which is both computationally expensive and time-consuming to collect. 
OSSN~\cite{zheng2024transformer}, a Transformer-based framework, unifies radio map estimation and site selection to reduce reliance on exhaustive candidate evaluations. While it improves planning efficiency, its performance remains limited by the availability of training data and a lack of real-world validation. 
Similarly, Dai and Zhang~\cite{dai2020propagation} take a hybrid approach by integrating machine learning with heuristic optimization, using a dataset of over $0.76$ million measurements to predict signal strength without explicit propagation modeling. 

However, existing AI/ML-based methods require extensive labeled datasets and incur substantial data collection and training costs. 
More importantly, they tend to overlook the sim-to-real discrepancy introduced by inaccurate or incomplete modeling of physical environments, which hinders generalization in real-world deployments.

\section{Conclusion}
In this paper, we propose a new automatic network planning framework (\emph{AutoPlan}) by leveraging digital radio twin techniques.
We derive the DRT by finetuning the parameters of building materials to reduce the sim-to-real discrepancy based on crowdsource real-world user data.
Leveraging the DRT, we design a Bayesian optimization-based algorithm to efficiently search for optimal base station parameters.
Using the field measurement from Husker-Net, we extensively evaluate \emph{AutoPlan} under various deployment scenarios, in terms of coverage, capacity, and efficiency.

\section*{Acknowledgment}
This work is partially supported by the US National Science Foundation under Grant No. 2321699, and No. 2426481. 

This material is based upon work supported by the Federal Motor Carrier Safety Administration under grant number: FM-MHP-0888.  Any opinions, findings, and conclusions or recommendations expressed in this publication are those of the author(s) and do not necessarily reflect the view of the Federal Motor Carrier Safety Administration and/or the U.S. Department of Transportation.

\bibliographystyle{IEEEtran}
\bibliography{ref/reference, ref/qiang, ref/ref}

\end{document}